\input{aipcheck}

\documentclass[
    ,final            
  ]
  {aipproc}

\layoutstyle{6x9}

\begin{document}

\title{Low Mass X-ray Binaries and Black Holes in Globular Clusters}


\author{Arunav Kundu}{
  address={Physics \& Astronomy Department, Michigan State University, East Lansing, MI 48824, USA}
}

\author{Thomas J. Maccarone}{
  address={School of Physics \& Astronomy, University of Southampton, 
Southampton, Hampshire SO17 1BJ}
}

\author{Stephen E. Zepf}{
  address={Physics \& Astronomy Department, Michigan State University, East Lansing, MI 48824, USA}
}

\author{I. Chun Shih}{
  address={Physics \& Astronomy Department, Michigan State University, East Lansing, MI 48824, USA}
}

\author{Katherine L. Rhode}{
  address={Department of Astronomy, Indiana University,
Bloomington, IN 47405}
}

\begin{abstract}
 Studies of nearby galaxies reveal that roughly half of their low mass X-ray binary (LMXB) populations are associated with globular clusters (GCs). We have established that the LMXB hosting frequency is correlated to various GC properties such as mass and metallicity. While the X-ray luminosities of a few of the brightest LMXBs in GCs are consistent with the accreting object being a black hole (BH), the only definitive way to distinguish between a black hole and multiple superposed sources in a GC is to detect variability. We have discovered just such a variable 4$\times$10$^{39}$ erg s$^{-1}$ black hole X-ray binary in a low metallicity globular cluster in the halo of NGC~4472. The change in the X-ray spectrum between the bright and faint epochs suggests that the luminosity variation is due to eclipsing by a warped accretion disk. The optical spectrum of this source also reveals strong, broad,  [O~III]~$\lambda$5007 and [O~III]~$\lambda$4959 emission. An analysis of the X-ray spectrum suggests that the [O~III] lines are produced by the photoionization of a wind driven by a stellar mass black hole accreting mass at or above its Eddington luminosity. As it is dynamically implausible to form an accreting stellar mass BH system in a GC with an intermediate mass BH it appears that this massive globular cluster does {\it not} harbor an intermediate mass BH. The inferred mass of this BH falls well below the extrapolation of the well known M$_{BH}$-$\sigma$ and M$_{BH}$-M$_{Stellar}$ relations to this GC.  Therefore our analysis suggests that not all old, metal poor stellar systems form black holes consistent with these relations, which have been established for much more massive stellar systems. 

 \end{abstract}

\maketitle


\section{The Globular Cluster - Low Mass X-ray Binary Connection }

The dense inner regions of globular clusters (GCs) are fertile grounds for the study of the relatively rare evolutionary end products of high mass stars, and exotic binary systems associated with such objects. For example, it has long been known that low mass X-ray binaries (LMXBs), which are close binary systems with an accreting neutron star (NS) or black hole (BH), are a few hundred times more abundant in globular clusters of the Milky Way than in the field. This is attributed to the efficient formation of LMXBs in globular clusters due to dynamical interactions in the core.

 Elliptical and S0 galaxies are ideal for detailed studies of the LMXB-GC link as they are particularly abundant in globular clusters, with at least an order of magnitude higher GC density than the Galaxy. The identification of LMXBs with these simple stellar systems that have well defined properties such as metallicity and age provides a unique opportunity to probe the effects of these parameters on LMXB formation and evolution. High resolution Chandra and XMM X-ray images of nearby ellipticals and S0s have resolved large numbers of point sources,  confirming a long-standing suggestion that the hard X-ray emission in many of these galaxies is predominantly from X-ray binaries. Moreover, most of the bright, L$_X$$\geq$10$^{37}$ erg s$^{-1}$ sources seen in typical observations of such galaxies must be LMXBs since they generally have stellar populations that are at least a few Gyrs old.

	Combined optical and X-ray studies reveal that roughly half of the LMXBs in early type galaxies are located in GCs (e.g. Kundu et al. 2002, 2007). These studies have established that LMXBs are preferentially found in the most luminous and red (metal-rich) GCs. Statistical tests reveal a marginal tendency of LMXBs to favor GCs with smaller half light radius, and no convincing correlation with galactocentric distance. 

	The underlying reasons for these correlations provide a window into both the dynamics of GCs and the physics of LMXBs. Since luminous clusters are known to be denser than less luminous ones and obviously have more stars, the higher rate of LMXB formation in these clusters due to the consequently higher dynamical interaction rate is not surprising. On the other hand there are no obvious dynamical reasons for the three times larger rate of LMXBs in red, metal-rich, globular clusters as compared to the metal-poor ones. LMXBs are similarly found preferentially in the metal-rich Galactic GCs but in the past this was often dismissed as the consequence of the location of these clusters in the bulge, where the dynamical evolution of GCs may be accelerated. The lack of correlation of the LMXBs with galactocentric distance in the large GC samples of ellipticals (Kundu et al. 2002, 2007) argues against this possibility and establishes a primary correlation with GC color. There have been some theoretical attempts to explain the metallicity effect (e.g. Maccarone, Kundu, \& Zepf 2004).

A small fraction of the LMXBs in early type galaxies, nearly all of which are in GCs, are brighter than the Eddington luminosity for accretion on to a neutron star and are hence candidate black holes. Irwin (2006) analyzed archival observations of some of these galaxies and concluded that these black hole candidates showed surprisingly little variation over the timescale of a few years. However, the various correlations between host GC properties and the rate of LMXBs in clusters indicates that it is likely that some GCs host multiple bright LMXBs (Kundu et al. 2007). Thus, temporal variability in the X-ray flux over a short time scale provides a definitive way to identify a BH in a globular cluster. We have discovered just such a black hole in a globular cluster in the nearby giant Virgo cluster elliptical galaxy NGC~4472. 

\section{A Black Hole in the NGC 4472 Globular Cluster RZ 2109}

\begin{figure}
  \includegraphics[width=8cm]{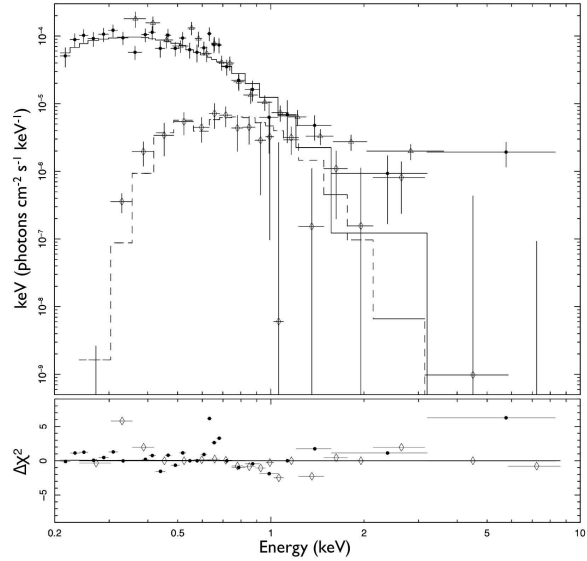}
  \caption{XMM-Newton EPIC spectrum and $\Delta$$\chi^{2}$ residuals for an absorbed MCD model fit to the spectrum of RZ 2109 the black hole in a NGC~4472 globular cluster during the bright (solid circle, solid line) and faint (open circles, dashed lines) epochs (from Shih et al. 2008). The open triangles indicate archival Chandra observations (that predate the XMM observations by about 4 years) which are in excellent agreement with the XMM spectrum of the bright phase.}
\end{figure}

We have discovered a bright 4$\times$10$^{39}$ erg s$^{-1}$ black hole X-ray binary in the high mass (7.5$\times$10$^{5}$M$_{\odot}$), spectroscopically identified globular cluster RZ 2109 in the halo of NGC 4472 (Maccarone et al. 2007). The observed count rate of this source varied by a factor of $\approx$7 over a few hours. The high X-ray luminosity, about 10 times L$_{Edd}$ for a NS, rules out any single object other than a black hole in the old stellar population of the GC. This observation provides the first unambiguous detection of a black hole X-ray binary in any globular cluster.

	In the high count rate interval (Fig. 1) the spectrum is well fit by a disk black model with an inferred inner disk radius of about 4,400 km and foreground Galactic absorption of 1.67$\times$10$^{20}$ H atoms cm$^{-2}$. The variation in the X-ray flux can mostly be attributed to the soft X-ray bands suggesting increased absorption (Fig 2). In fact the spectrum of the low count interval is consistent with having the same underlying continuum as the bright phase, but an increased neutral Hydrogen column density of $\approx$3$\times$10$^{21}$ H atoms cm$^{-2}$ (Fig 1). The source varies by about a factor of 10 below 0.7 keV and is consistent with being constant above 0.7 keV. A Chandra observation predating the XMM observations by 3.5 years also reveals a spectrum consistent with the bright phase (Shih et al. 2008). This luminous source was also detected by ROSAT 10 years earlier, indicating that this BH source has been in a bright X-ray phase  for at least a decade.

The ingress time of $\approx$10 ks provides a key constraint on the nature of source. A straightforward calculation combining the ingress time with Kepler's law rules out occultation by a Roche lobe filling secondary companion, both because it leads to an unphysically long orbital period and because it indicates a semi-major axis that is of the scale of the host globular cluster (Shih et al. 2008). The more likely scenario is that the X-ray source is being quasi-periodically obscured by a tilted accretion disk precessing around it on a long period. It is known that accretion disks can be tilted by the tidal force of the companion star, and/or warped by the effects of radiation pressure on an accretion disc. The precession period of such a warped disk is much longer than the orbital period of the binary system. The parameters of the RZ 2109 BH are consistent with the observations of other warped systems (Shih et al. 2008).

\begin{figure}
  \includegraphics[width=7cm]{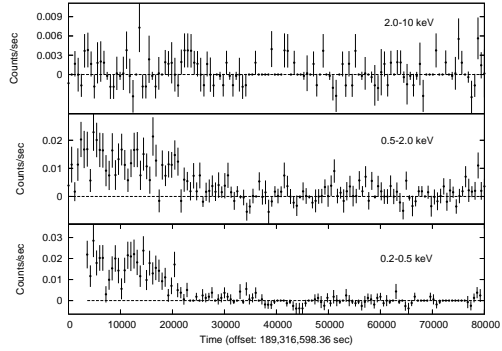}
  \caption{The XMM-Newton light curves of RZ 2109 in three X-ray bands. Most of the variation appears to be in the soft bands suggesting foreground absorption.}
\end{figure}

The X-ray observations in themselves do not place strong constraints on the mass of the black hole and allow for either a roughly 10M$_{\odot}$ stellar mass BH or a $\geq$400M$_{\odot}$ intermediate mass BH. There is much interest in BHs that reside in GCs because some dynamical models predict the formation of intermediate mass BHs due to mergers in the core. The extrapolation of the M$_{BH}$-$\sigma$ relation also suggests that intermediate mass BHs may reside in globular clusters. However other theoretical studies of stellar interactions in globular clusters suggest that most BHs should be ejected from GCs due to dynamical interactions in the core, leaving one stellar mass BH or a BH binary in the core. The study of the optical spectrum of RZ 2109 described below provides an interesting constraint on  both the mass of the black hole in this GC and the larger discussion of intermediate mass black holes in globular clusters.

\section{Clues to the Nature of the RZ 2109 Black Hole from X-Ray and Optical Spectroscopy }

\begin{figure}
  \includegraphics[angle=270,width=9.5cm]{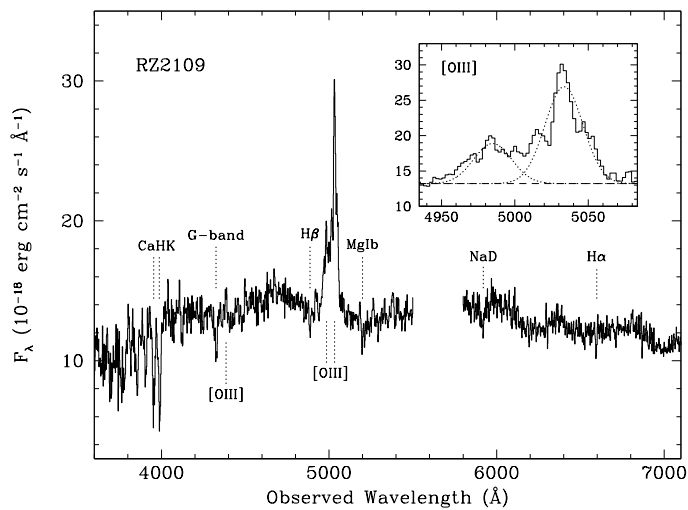}
  \caption{A plot of $F_{\lambda}$ vs. $\lambda$ for RZ 2109, the NGC~4472 
globular cluster with a black hole system from Zepf et al. 2008. This figure shows the remarkably strong and broad [O~III]$\lambda\lambda$4959, 5007 [O~III] emission lines. The large [O~III]/H$\beta$ ratio indicates that the [O~III] lines are not produced directly by shocks, but are likely photoionized. The upper inset figure shows the [O~III]$\lambda\lambda$4959, 5007
region in more detail and demonstrates the extraordinary width
of these lines. The gap around 5600 \AA\ is due
to the dichroic in the Keck-LRIS spectrograph. }
\end{figure}

The optical globular cluster candidate RZ 2109 in NGC 4472 was confirmed to be a globular cluster by two VLT spectroscopic observations. Analysis of the absorption lines associated with the spectrum of the old stellar population of this globular cluster indicated a consistent radial velocity of 1477$\pm$7 km s$^{-1}$ and 1460$\pm$19 km s$^{-1}$ in the two measurements (Zepf et al. 2007). This unambiguously identifies the object as a member of the NGC 4472 globular cluster system, which has a mean radial velocity of 1018 km s$^{-1}$ and a velocity dispersion of $\sim$300 km s$^{-1}$.

More interestingly an inspection of both spectra revealed a broad emission line at a wavelength of 5031.2$\pm$0.3 \AA. This corresponds precisely to [O~III]~$\lambda$5007 redshifted by the radial velocity of the cluster. The fact that  it was observed in the same place in two spectra confirmed this interpretation (Zepf et al. 2007). A subsequent Keck observation with broader wavelength coverage (Fig 3) not only shows wide [O~III]~$\lambda$5007 emission but [O~III]~$\lambda$4959 and possibly [O~III]~$\lambda$4363 emission lines at the radial velocity of the globular cluster confirming beyond doubt that these lines are associated with the globular cluster RZ 2109 which hosts a black hole X-ray binary. 

There spectrum plotted in Fig 3 has two remarkable features. It reveals strong [O~III] emission with surprisingly little emission associated with other lines. The [O~III] line is also remarkably broad, so broad in fact that the $\lambda$4959 and $\lambda$5007 are blended. Using standard de-blending techniques we found that the FWHM of the line is 33 \AA, which corresponds to a velocity FWHM of approximately 2000 km s$^{-1}$. It is worth noting that emission lines are rarely observed in old globular clusters such as RZ 2109. Of the hundreds of GCs studied in many nearby galaxies only about half a dozen show evidence of emission lines. These lines are usually tens of km s$^{-1}$ wide, orders of magnitude narrower than the RZ 2109 [O~III] emission line, and are associated with planetary nebulae. We conclude in Zepf et al. (2007, 2008) that the broad [O~III]~$\lambda$5007 line observed in RZ 2109, as broad as almost any such line observed anywhere, is associated with the black hole in this globular cluster.

The optical spectrum provides important insights into the nature of the emission process, and the mass of the black hole. In Zepf (2007) we discussed three possible mechanisms for producing the broad [O~III] line. The line can be produced by shocks from the interaction of a strong, high velocity wind driven from the black hole with the interstellar gas in the cluster. Another possibility is that the strong wind is photoionized by the powerful central system. Extensive theoretical work has found that such strong winds are driven only by systems with mass accretion rates similar to, or larger than, the Eddington limit. Thus the 4$\times$10$^{39}$ erg s$^{-1}$ luminosity of the X-ray source would imply a stellar mass ($\approx$10M$_{\odot}$) BH if a stellar wind were confirmed.  Alternately the broad linewidth can be caused by the gravitational motion of photoionized material very close to the BH. Such a scenario would indicate an intermediate mass black hole, in order to account for the high velocities at sufficient distances to allow [O~III] emission.

It is immediately apparent from Fig 3 that the H$\beta$ emission is not strong; The Balmer lines are all seen in absorption. The large [O~III]~$\lambda$5007/H$\beta$ emission line ratio of about 30, after accounting for H$\beta$ stellar absorption lines, rules out shocks as the primarily mechanism for producing the [O~III] line (Zepf et al. 2008). The L([O~III]~$\lambda$5007)~=~1.4$\times$10$^{37}$ erg s$^{-1}$ luminosity of the [O~III] line coupled with the broad $\pm$1500 km s$^{-1}$ wings of the lines also argue strongly against the intermediate mass black hole scenario. We show in Zepf et al. (2008) that the maximum possible [O~III] flux from a gas at the critical density of [O~III]~$\lambda$5007 that completely fills the volume required for the circular velocity around an intermediate mass black hole to match the observed [O~III]~$\lambda$5007 line width is almost 8 orders of magnitude less than the observed [O~III] line luminosity. Thus we conclude that the [O~III]~$\lambda$5007 emission is caused by the photoionization of material driven across the cluster by winds from the RZ 2109 black hole, and that the object is a stellar mass black hole.

Dynamical studies of GCs suggest that it is highly unlikely that stellar mass black hole binary systems could form in a cluster with an intermediate mass black hole at its center. A GC with an intermediate mass BH is expected to eject stellar mass black holes from the host cluster. Thus, our analysis indicates that this particular globular cluster only harbors a stellar mass black hole, and does not host an intermediate mass BH. This is particularly interesting because it has often been suggested based on the extrapolation of the well known M$_{BH}$-$\sigma$ and M$_{BH}$-M$_{Stellar}$ relationship for BHs that the hitherto elusive intermediate mass BHs may be found in GCs. We show in Zepf et al. (2008) that the BH in RZ 2109 falls far off these relations which predict a BH that is roughly two orders of magnitude more massive than the one we have discovered in RZ 2109. Our analysis reveals that not all old, metal poor stellar systems form black holes consistent with the M$_{BH}$-$\sigma$ and M$_{BH}$-M$_{Stellar}$ relations found for more massive galaxies. It also indicates that at least in this fairly massive globular cluster stellar dynamics did not lead to the coalescence of massive objects at the center of this globular cluster into and intermediate mass black hole.
 

\begin{theacknowledgments}
 AK gratefully acknowledges support from Chandra grant SAO AR7-8010X and NASA LTSA grant NAG5-12975 for this research.
\end{theacknowledgments}

\end{document}